\documentclass[conference]{IEEEtran}
\IEEEoverridecommandlockouts
\usepackage{graphicx}
\usepackage{booktabs}
\usepackage{hyperref}
\usepackage{tabularx}
\usepackage{array}
\usepackage{cite}
\usepackage{tikz} % For flowcharts
\usepackage{amsmath}
\usepackage{xcolor}
\usetikzlibrary{calc, positioning, arrows, shapes.geometric, fit,backgrounds}
\usepackage{listings}
\usepackage[compatibility=false]{caption}
\usepackage{xcolor}
\usepackage{braket}
\usepackage{graphicx}
\usepackage{subcaption}
\usepackage{multirow}
\usepackage{dblfloatfix}
\usepackage[utf8]{inputenc}
\usepackage{amssymb}
\definecolor{lightgray}{gray}{0.95}

\usepackage{algorithm}
\usepackage{algpseudocode}
\usepackage{float}    % 讓 algorithm 不被拆開
\usepackage{ragged2e} % 讓演算法內容不強制對齊

\begin{document}

\title{Quantum-Enhanced Reinforcement Learning with LSTM Forecasting Signals for Optimizing Fintech Trading Decisions \thanks{The views expressed in this article are those of the authors and do not represent the views of Wells Fargo. This article is for informational purposes only. Nothing contained in this article should be construed as investment advice. Wells Fargo makes no express or implied warranties and expressly disclaims all legal, tax, and accounting implications related to this article.}}

% \author{
% \IEEEauthorblockN{1st Given Name Surname}
% \IEEEauthorblockA{Dept. Name of Organization\\
% Name of Organization\\
% City, Country\\
% Email: example@domain.com}
% \and
% \IEEEauthorblockN{2nd Given Name Surname}
% \IEEEauthorblockA{Dept. Name of Organization\\
% Name of Organization\\
% City, Country\\
% Email: example2@domain.com}
% }
\author{
    \IEEEauthorblockN{Yen-Ku Liu\IEEEauthorrefmark{1}, Yun-Huei Pan\IEEEauthorrefmark{1}, Pei-Fan Lu\IEEEauthorrefmark{1}, Yun-Cheng Tsai\IEEEauthorrefmark{1}, Samuel Yen-Chi Chen\IEEEauthorrefmark{2}}
    \IEEEauthorblockA{\IEEEauthorrefmark{1}Department of Technology Application and Human Resource Development, National Taiwan Normal University, Taipei, Taiwan\\ Email: pecu@ntnu.edu.tw}
    \IEEEauthorblockA{\IEEEauthorrefmark{2}Wells Fargo, New York, USA\\ Email: yen-chi.chen@wellsfargo.com}
}

\maketitle

\begin{abstract}
Financial trading environments are characterized by high volatility, numerous macroeconomic signals, and dynamically shifting market regimes, where traditional reinforcement learning methods often fail to deliver breakthrough performance. In this study, we design a reinforcement learning framework tailored for financial systems by integrating quantum circuits. We compare (1) the performance of classical A3C versus quantum A3C algorithms, and (2) the impact of incorporating LSTM-based predictions of the following week's economic trends on learning outcomes. The experimental framework adopts a custom Gymnasium-compatible trading environment, simulating discrete trading actions and evaluating rewards based on portfolio feedback. Experimental results show that quantum models—especially when combined with predictive signals—demonstrate superior performance and stability under noisy financial conditions, even with shallow quantum circuit depth.
\end{abstract}


\section{Introduction}
Financial markets are inherently complex, stochastic, and often exhibit sudden regime shifts driven by macroeconomic developments, geopolitical events, and evolving investor sentiment \cite{cont2001empirical}. Traditional algorithmic trading strategies—ranging from simple technical-rule systems to deep-learning-based predictors—struggle to adapt swiftly to such non-stationary environments \cite{zhang2019deep}. Reinforcement learning (RL), by framing trading as a sequential decision-making problem, has emerged as a robust framework for learning adaptive strategies directly from market data \cite{moody1998performance, deng2016deep}.

Among RL algorithms, Asynchronous Advantage Actor-Critic (A3C) \cite{mnih2016asynchronous} stands out for its sample efficiency and stability. Multiple parallel agents explore concurrently, sharing gradients with a global network to accelerate learning and mitigate the high variance in policy-gradient methods. However, classical A3C agents may lack representational capacity for intricate, high-dimensional financial time series dependencies \cite{henderson2018deep}.

Quantum machine learning (QML) offers powerful tools to address these challenges. Quantum Neural Networks (QNNs) and Variational Quantum Algorithms (VQAs) leverage superposition and entanglement for nonlinear transformations, embedding data into large Hilbert spaces to distinguish subtle patterns \cite{biamonte2017quantum, Schuld2015, Cerezo2021}. These quantum feature maps reveal correlations that are overlooked by classical methods, showing promise in applications such as blockchain analytics \cite{Havlicek2019, tsai2025quantumfeatureoptimizationenhanced}. Quantum algorithms also demonstrate advantages in high-dimensional or noisy datasets \cite{Dunjko2018, Preskill2018, schuld2019quantum}.

Quantum computing enhances representational power via Variational Quantum Circuits (VQCs) in expansive Hilbert spaces \cite{biamonte2017quantum}. Quantum feature maps encode inputs into states, providing nonlinear embeddings similar to those in kernel methods \cite{schuld2019quantum}. Hybrid architectures improve classification on real hardware \cite{havlivcek2019supervised}. Recently, Chen et al. proposed a quantum-enhanced variant of the A3C algorithm and demonstrated its empirical advantages over classical baselines \cite{chen2023asynchronous}. Subsequent advancements include the integration of recurrent policies \cite{chen2024efficient} and the application of differentiable quantum architecture search within the A3C framework \cite{chen2024differentiable}. However, the application of quantum RL to real-world financial tasks remains largely underexplored.
% Recently, Chen et al. proposed the quantum-enhanced version of A3C and demonstrates empirical advantages over classical counterparts \cite{chen2023asynchronous}. Further advancements of A3C including recurrent policies \cite{chen2024efficient} and differentiable quantum architecture search in A3C \cite{chen2024differentiable}. Yet, quantum RL applications to real-world financial tasks remain largely unexplored.

In this work, we introduce a \emph{Fintech Quantum A3C} agent for weekly S\&P\,500 trading, augmented with LSTM-based macroeconomic forecasts. Our key contributions include a head-to-head comparison of classical A3C and hybrid quantum-classical A3C on S\&P 500 trading, highlighting differences in representational capacity and learning dynamics. We also integrate one-week-ahead LSTM forecasts of macroeconomic and price features into both agents, studying how these anticipatory signals affect convergence speed, stability, and overall performance. Finally, we analyze how Quantum A3C—with and without forecasts—modifies trading behavior (entry/exit frequency, holding durations) and risk-adjusted metrics (Sharpe, Calmar ratios, drawdowns), demonstrating the advantages of the quantum-enhanced agent over its classical counterpart.

\section{Methodology}
\subsection{Dataset Description}
We focus on the period from January 2022 to April 2024, using S\&P 500 data, which exhibited high volatility relative to the past decade, making it an ideal time for testing models in turbulent markets. The dataset incorporates various macroeconomic and market indicators for a realistic trading setup, including VIX (a measure of market sentiment and volatility), FEDFUNDS, DGS2, and DGS10 (federal funds rate and U.S. Treasury yields), as well as BAMLH0A0HYM2 (a high-yield bond spread indicating credit risk).

All features are z-score normalized for training stability and scale comparability. To prevent data leakage, each point uses only current values at time $t$, maintaining causal consistency in the RL framework.

Predictive versions augment the dataset with LSTM forecasts of the following week's S\&P 500 direction, enhancing the agent's environmental perception.

\subsection{Environment Formalization}
We define the trading environment as a Markov Decision Process (MDP) represented by the tuple $\langle \mathcal{S}, \mathcal{A}, \mathcal{P}, r, \gamma \rangle$. The state space $\mathcal{S} \subseteq \mathbb{R}^n$ consists of $n$-dimensional vectors of normalized financial indicators at time $t$, such as interest rates, volatility indices, and stock prices. The discrete action space $\mathcal{A} = \{0, 1, 2\}$ corresponds to hold, buy, and sell actions. The transition function $\mathcal{P}$ is implicitly defined by historical market data sequences. The reward function $r: \mathcal{S} \times \mathcal{A} \to \mathbb{R}$ is given by
\[
r_t =
\begin{cases}
    p_t - p_{\text{buy}} - c \cdot p_t, & \text{if } a_t = 2 \text{ (sell)} \\
    0, & \text{otherwise},
\end{cases}
\]
where $p_t$ is the current S\&P 500 closing price, $p_{\text{buy}}$ is the buy price, and $c$ is the trade cost rate. The discount factor $\gamma$ is set to 0.9 for episodic evaluation.

At each step, the agent observes state $s_t$, selects action $a_t$, and receives reward $r_t$. The portfolio value updates as
\[
\text{Asset}_t = b_t + \mathbb{1}_{\text{position}=1} \cdot (p_t - p_{\text{buy}}),
\]
combining cash balance $b_t$ and unrealized profit. Episodes terminate at the end of the dataset.

\subsection{Fintech Quantum A3C Agent Architecture}

The Fintech Quantum A3C agent consists of two modular components: SP500TradingEnv, which defines the financial trading environment, and A3C Agent, which implements the Asynchronous Advantage Actor-Critic algorithm.

\begin{figure*}[htbp]
  \centering
  \includegraphics[width=\textwidth]{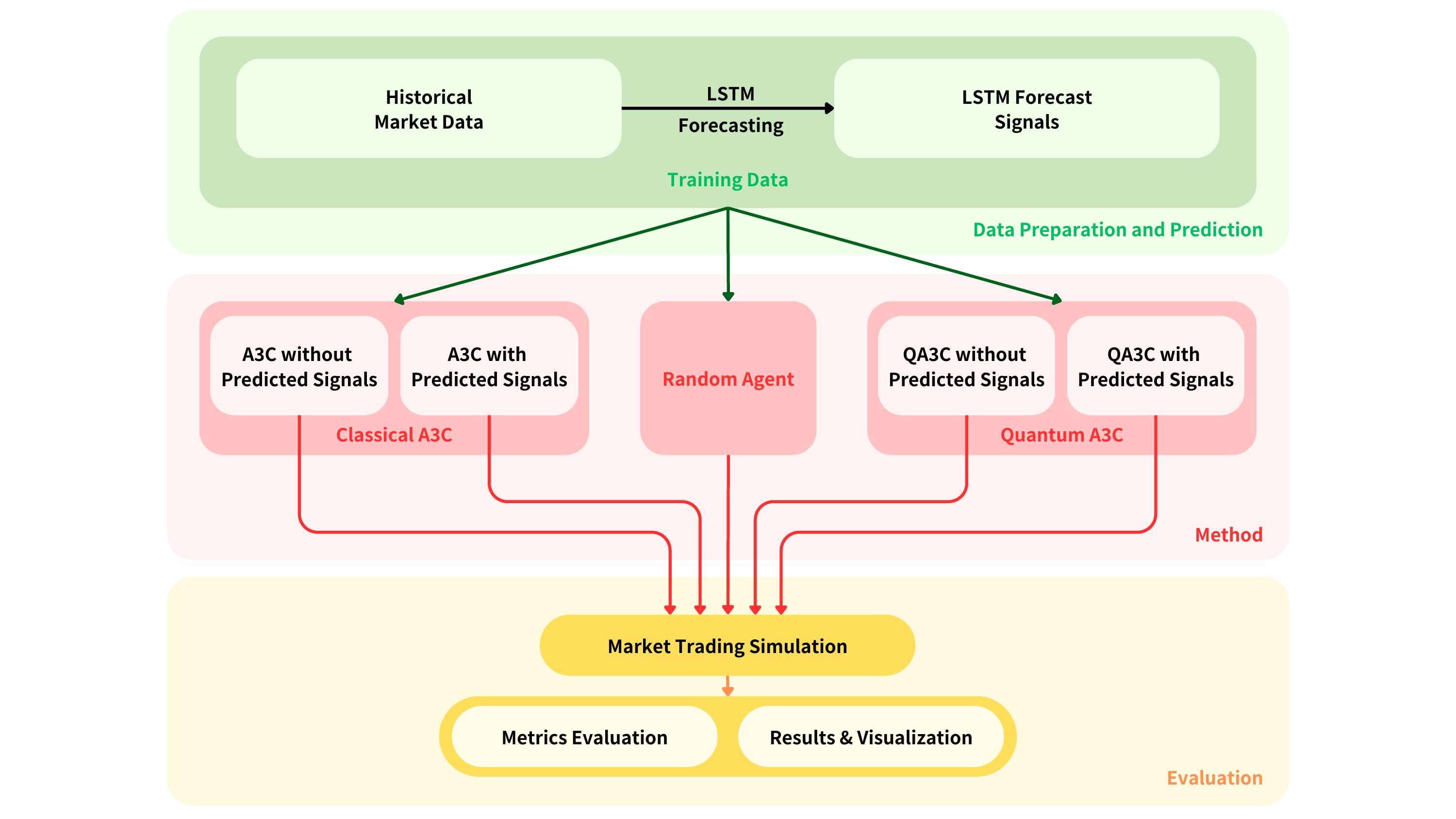}
  \caption{Experimental pipeline integrating LSTM-based prediction with Classical A3C, Quantum A3C, and Random trading agents.}
  \label{fig:experiment-flow}
\end{figure*}

The experimental framework includes four main stages: data preparation and LSTM-based forecasting, training, execution, and evaluation. Figure~\ref{fig:experiment-flow} illustrates the overall design.

\paragraph{Training Loop}  
Training initializes a global network $G(\theta)$ and shared optimizer, then spawns $N$ workers. Each worker $i$ instantiates its own SP500TradingEnv and local network $W_i(\theta_i \leftarrow \theta)$, repeating until the global episode count reaches $MAX\_EP$. It collects trajectories $\{(s_t, a_t, r_t)\}_{t=1}^K$ using policy $\pi(a_t \mid s_t; \theta_i)$. Every $K = UPDATE\_GLOBAL\_ITER$ steps or at terminal states, it computes discounted returns
\[
R_t = \sum_{k=0}^{K-1} \gamma^k r_{t+k} + \gamma^K V(s_{t+K}; \theta_i),
\]
gradients of the loss
\[
L = \frac{1}{2}(R_t - V(s_t))^2 - \log \pi(a_t \mid s_t)(R_t - V(s_t)),
\]
pushes $\nabla_{\theta_i} L$ to the global network via
\[
\theta \leftarrow \theta - \alpha \cdot \nabla_{\theta_i} L,
\]
And pulls updated parameters into $\theta_i$. Episode rewards are recorded, and global counters are incremented.

\paragraph{Quantum Encoding}

In the Fintech A3C framework, classical feedforward encoders in policy and value networks are replaced by a Variational Quantum Circuit (VQC). A learnable classical projection maps input state $s_t \in \mathbb{R}^d$ to latent vector $z_t \in \mathbb{R}^n$, which the VQC processes to yield a quantum-encoded output for subsequent linear layers in action-logit or state-value estimation.

Implemented via TorchVQC, a custom PennyLane-based PyTorch module, the circuit uses eight qubits, depth 2, and angle encoding with trainable $R_y$, $R_z$ rotations and entangling CNOT gates. This hybrid design leverages high-dimensional Hilbert space representations, serving as a nonlinear transformation to capture subtle patterns in financial features such as moving averages, volatility, and macroeconomic signals, thereby enhancing generalization in noisy markets. The forward pass for policy and value heads is summarized as
\[
% \begin{aligned}
% z_t &= \tanh(W_1 s_t), \\
% q_t &= \mathrm{VQC}(z_t), \\
% \pi(a_t \mid s_t) &= \mathrm{softmax}(W_2 \tanh(q_t)), \\
% V(s_t) &= W_4 \tanh(\mathrm{VQC}(W_3 s_t)),
% \end{aligned}
\begin{aligned}
\textbf{Policy head} \;(\pi): \quad
& z_t^{\pi} = \tanh\!\bigl(W_1 s_t + b_1\bigr), \\
& q_t^{\pi} = \mathrm{VQC}_{\pi}\!\bigl(z_t^{\pi}\bigr), \\
& \pi(a_t \mid s_t) = \mathrm{softmax}\!\bigl(W_2\,\tanh(q_t^{\pi}) + b_2\bigr), \\[6pt]
\textbf{Value head} \;(V): \quad
& z_t^{V} = \tanh\!\bigl(W_3 s_t + b_3\bigr), \\
& q_t^{V} = \mathrm{VQC}_{V}\!\bigl(z_t^{V}\bigr), \\
& V(s_t) = W_4 \,\tanh\!\bigl(q_t^{V}\bigr) + b_4.
\end{aligned}
\]
where $W_i$ are trainable classical layers and $\mathrm{VQC}$ is the parameterized quantum layer.

The learned policy is evaluated in SP500TradingEnv to generate asset trajectories, enabling comparisons with classical A3C and baselines, and demonstrating the feasibility of hybrid quantum-classical learning in financial tasks.

\paragraph{Evaluation \& Plotting}  
Post-training, the global policy $G(\theta)$ is executed once in SP500TradingEnv to generate the final asset history, which is visualized through full plotting for training progress and trading performance.

This modular design separates environment dynamics from learning algorithms, simplifying extension to Quantum A3C by substituting classical MLP encoders with Variational Quantum Circuits in policy/value networks.

\section{Experimental Results}
This section presents results from comparing quantum and classical reinforcement learning models, with and without LSTM-based predictive data, in the context of financial trading. Results are organized into training performance, trading behavior, and outcomes.

\begin{figure*}[htbp]
    \centering
    \begin{minipage}[t]{0.48\linewidth}
        \vspace{0pt}
        \begin{subfigure}{\linewidth}
            \includegraphics[width=\linewidth]{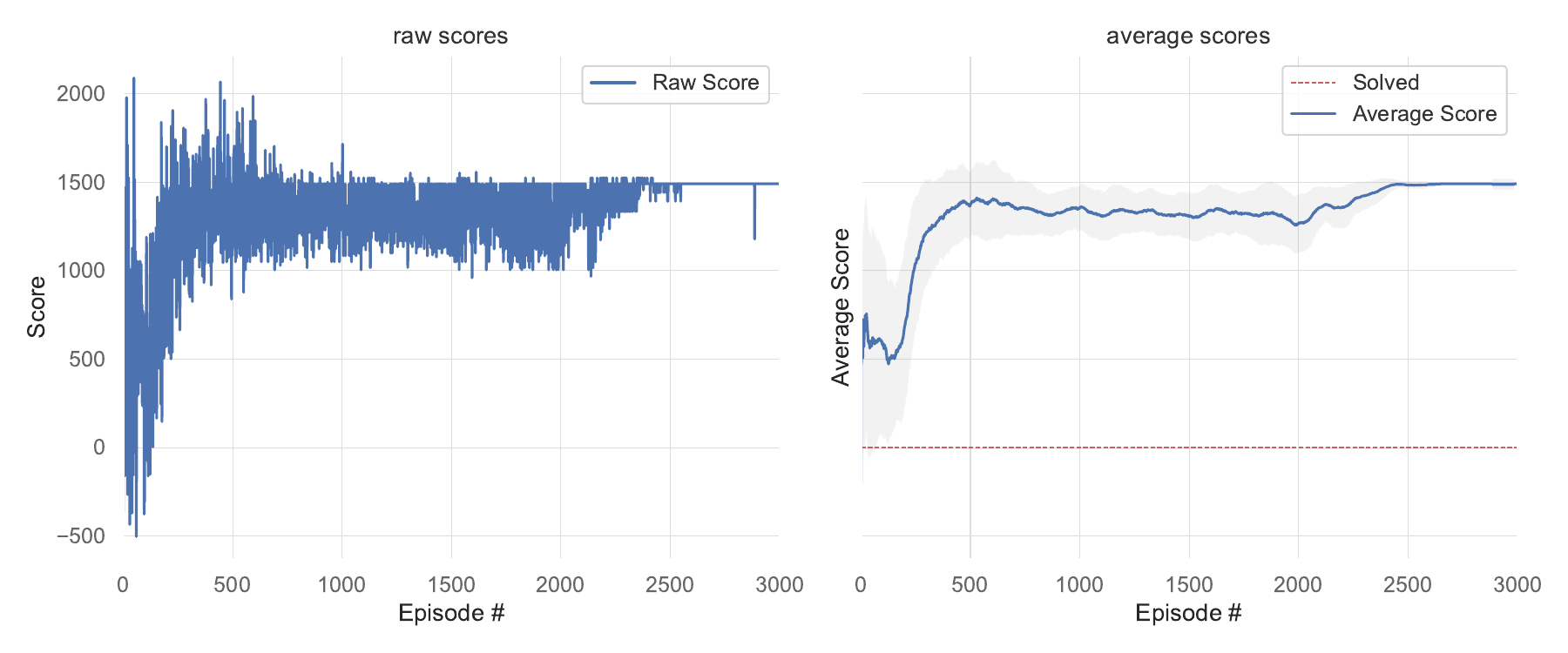}
            \caption{Classical A3C without forecasting}
        \end{subfigure}

        \vspace{0.3cm}

        \begin{subfigure}{\linewidth}
            \includegraphics[width=\linewidth]{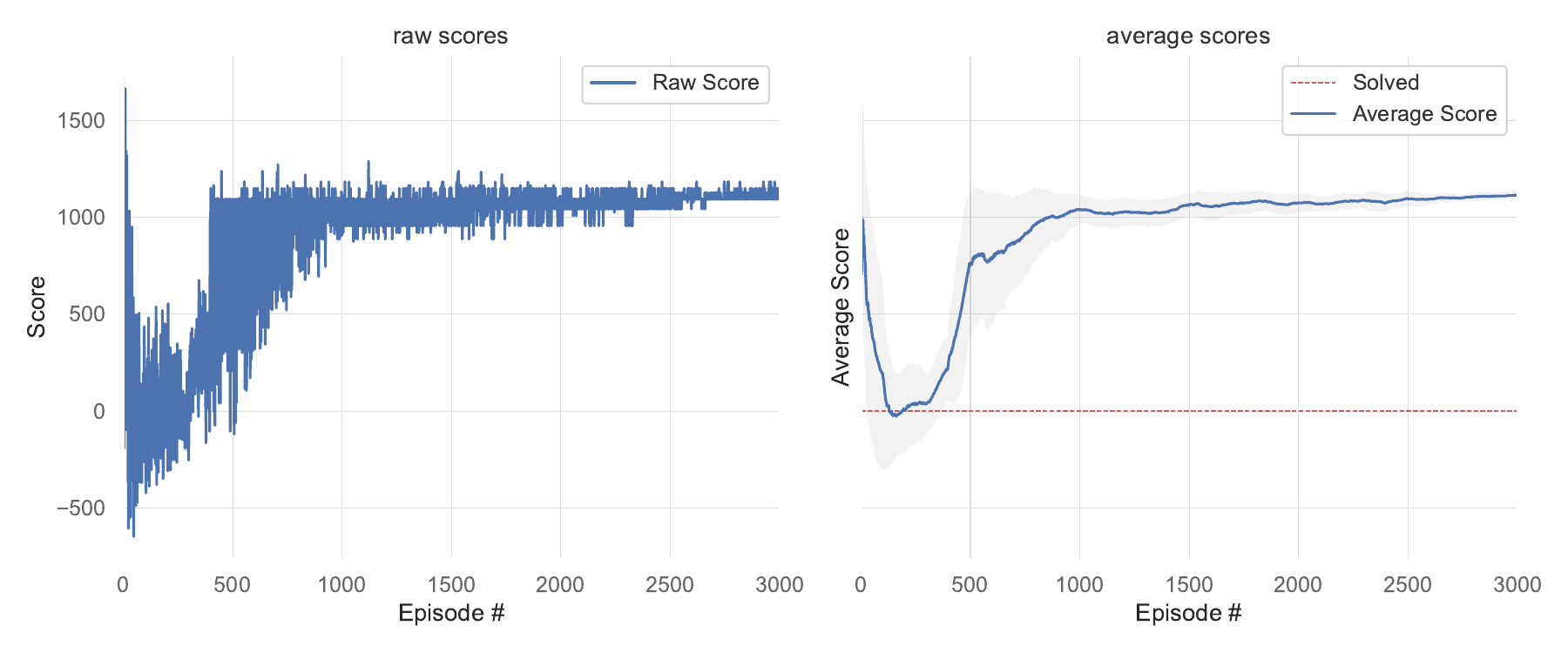}
            \caption{Classical A3C with LSTM forecasting}
        \end{subfigure}

        \vspace{0.3cm}

        \begin{subfigure}{\linewidth}
            \includegraphics[width=\linewidth]{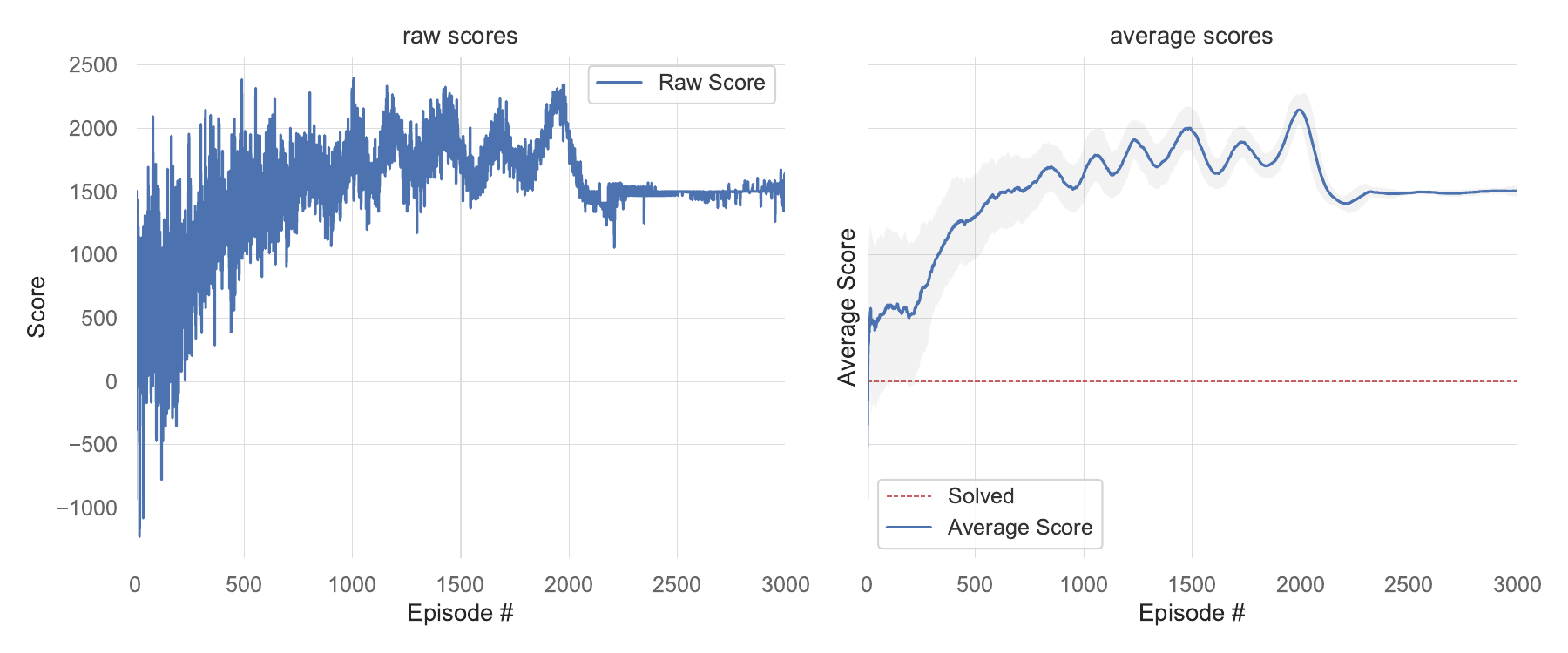}
            \caption{Quantum A3C without forecasting}
        \end{subfigure}

        \vspace{0.3cm}

        \begin{subfigure}{\linewidth}
            \includegraphics[width=\linewidth]{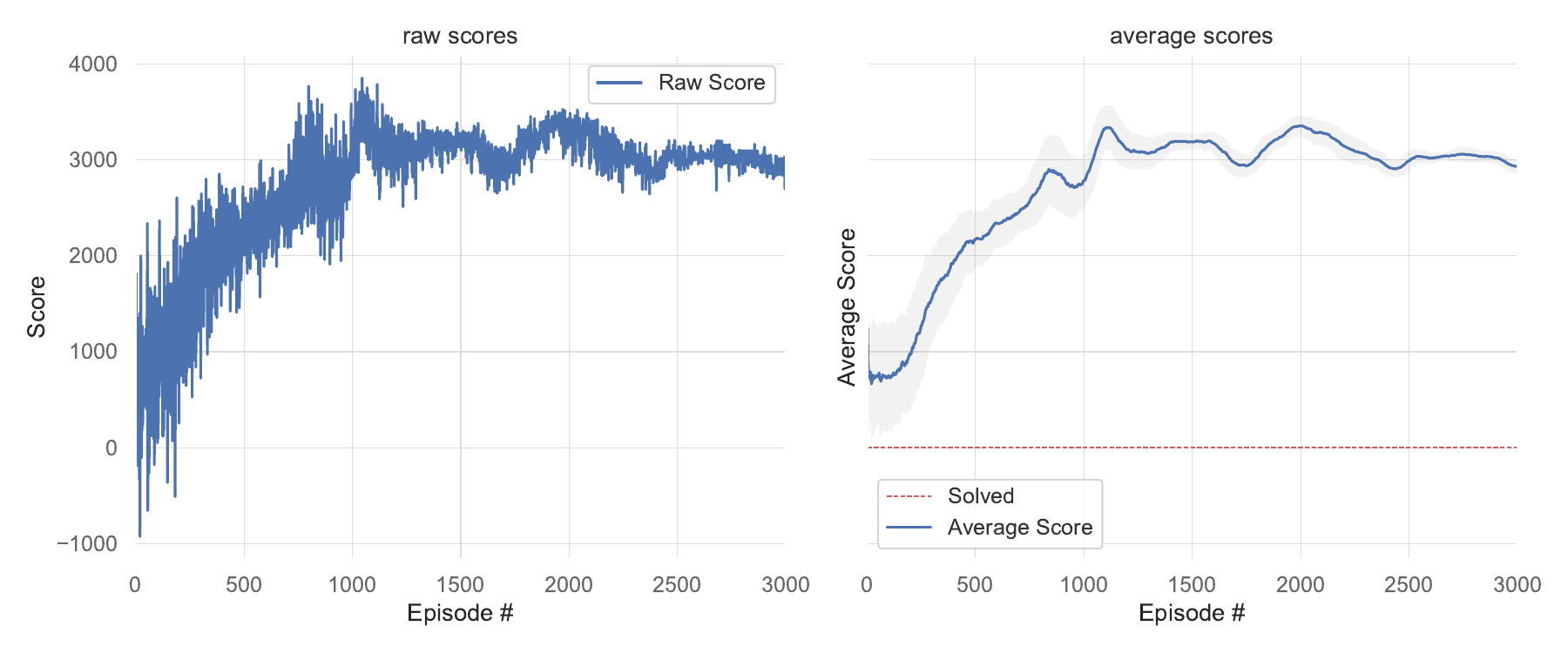}
            \caption{Quantum A3C with LSTM forecasting}
        \end{subfigure}

        \vspace{0.3cm}

        \begin{subfigure}{\linewidth}
            \includegraphics[width=\linewidth]{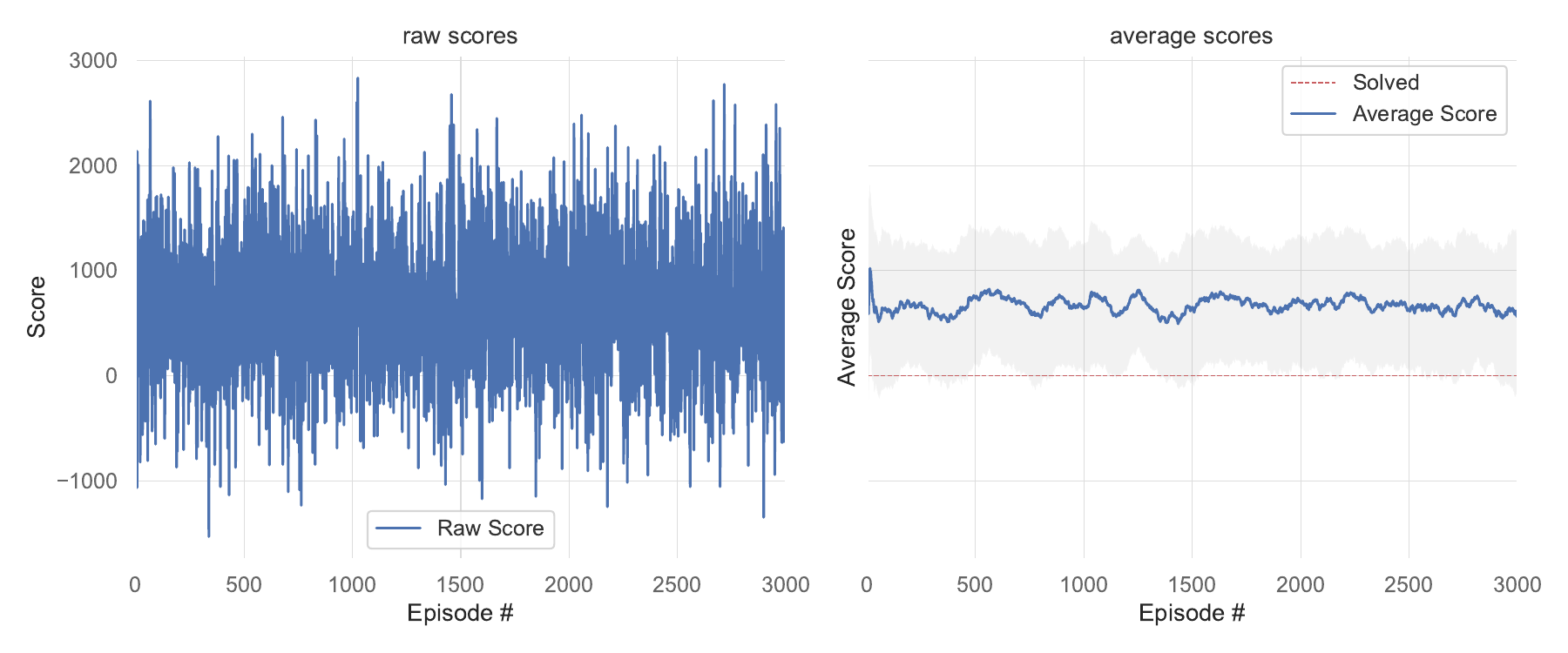}
            \caption{Random policy baseline}
        \end{subfigure}
    \end{minipage}
    \hfill
    \begin{minipage}[t]{0.48\linewidth}
        \vspace{0pt}
        \begin{subfigure}{\linewidth}
            \includegraphics[width=\linewidth]{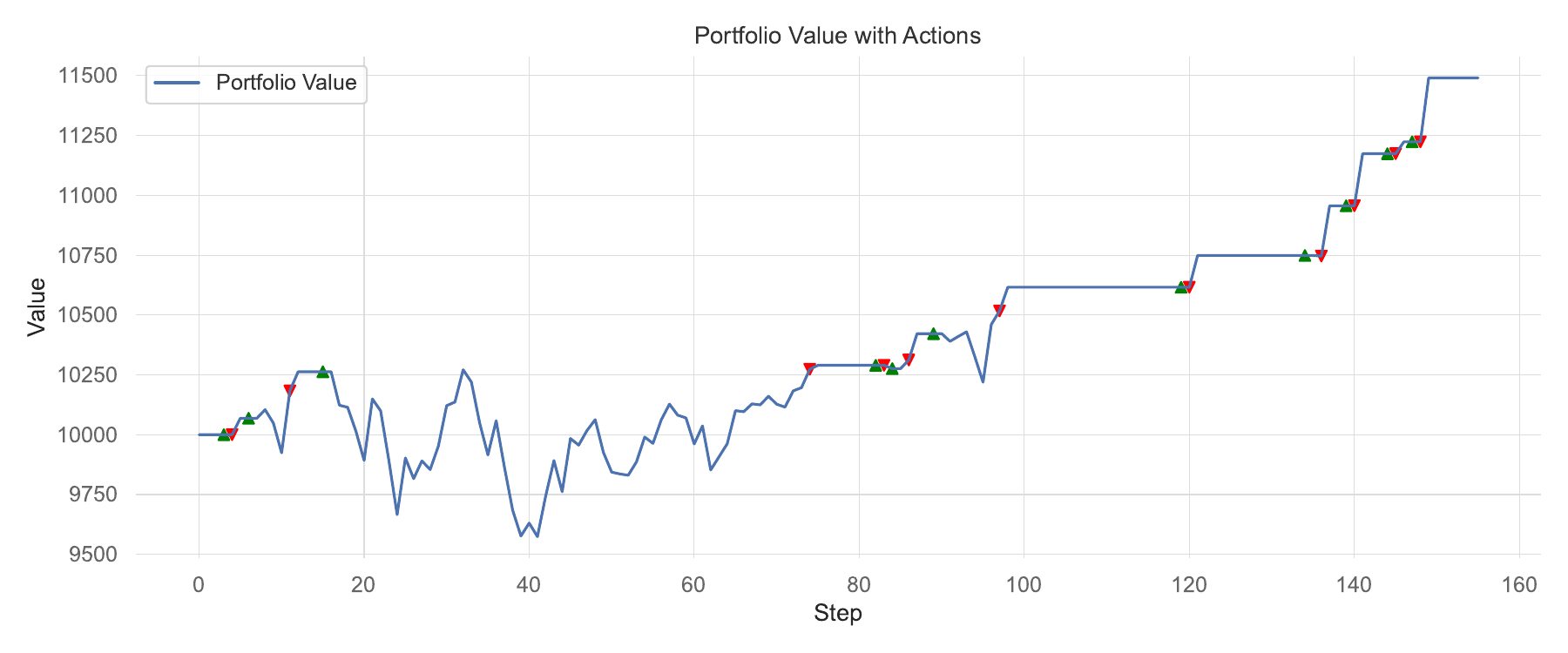}
            \caption{Classical A3C without forecasting}
        \end{subfigure}

        \vspace{0.3cm}

        \begin{subfigure}{\linewidth}
            \includegraphics[width=\linewidth]{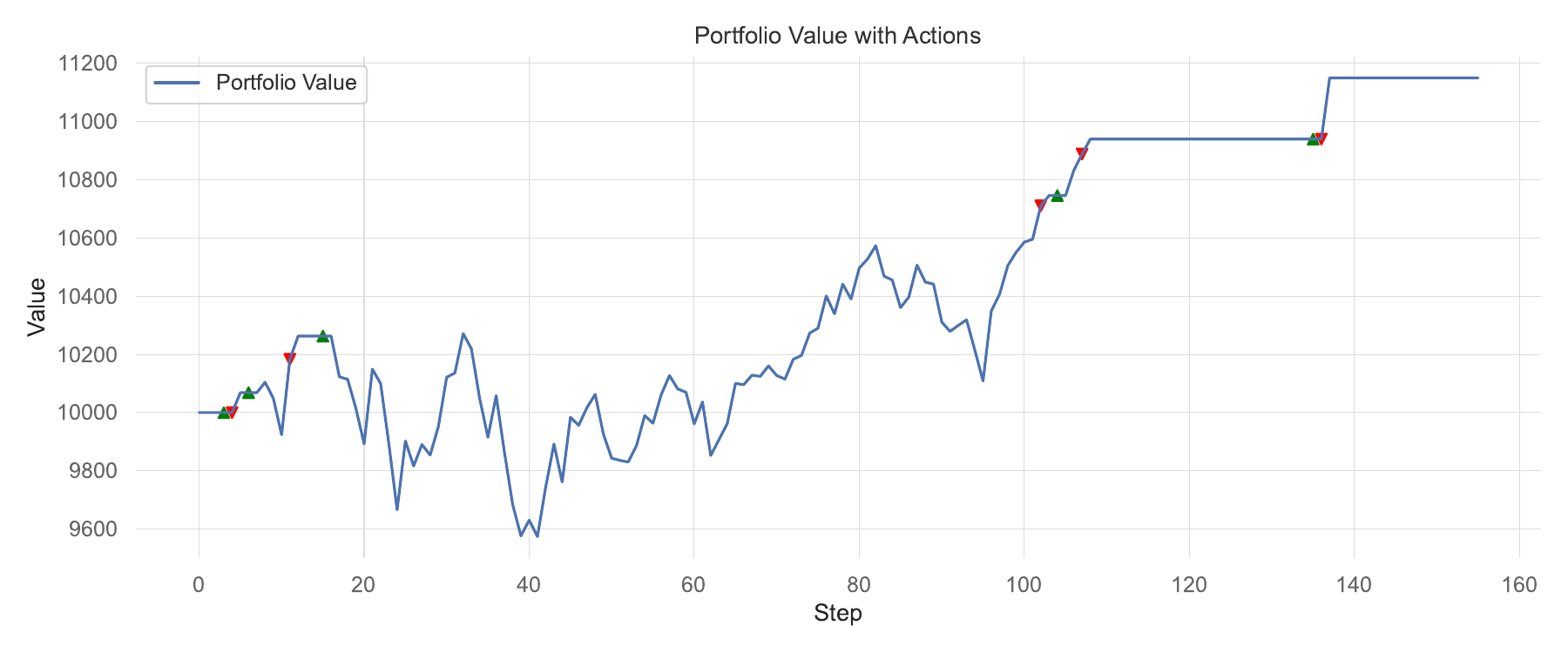}
            \caption{Classical A3C with LSTM forecasting}
        \end{subfigure}

        \vspace{0.3cm}

        \begin{subfigure}{\linewidth}
            \includegraphics[width=\linewidth]{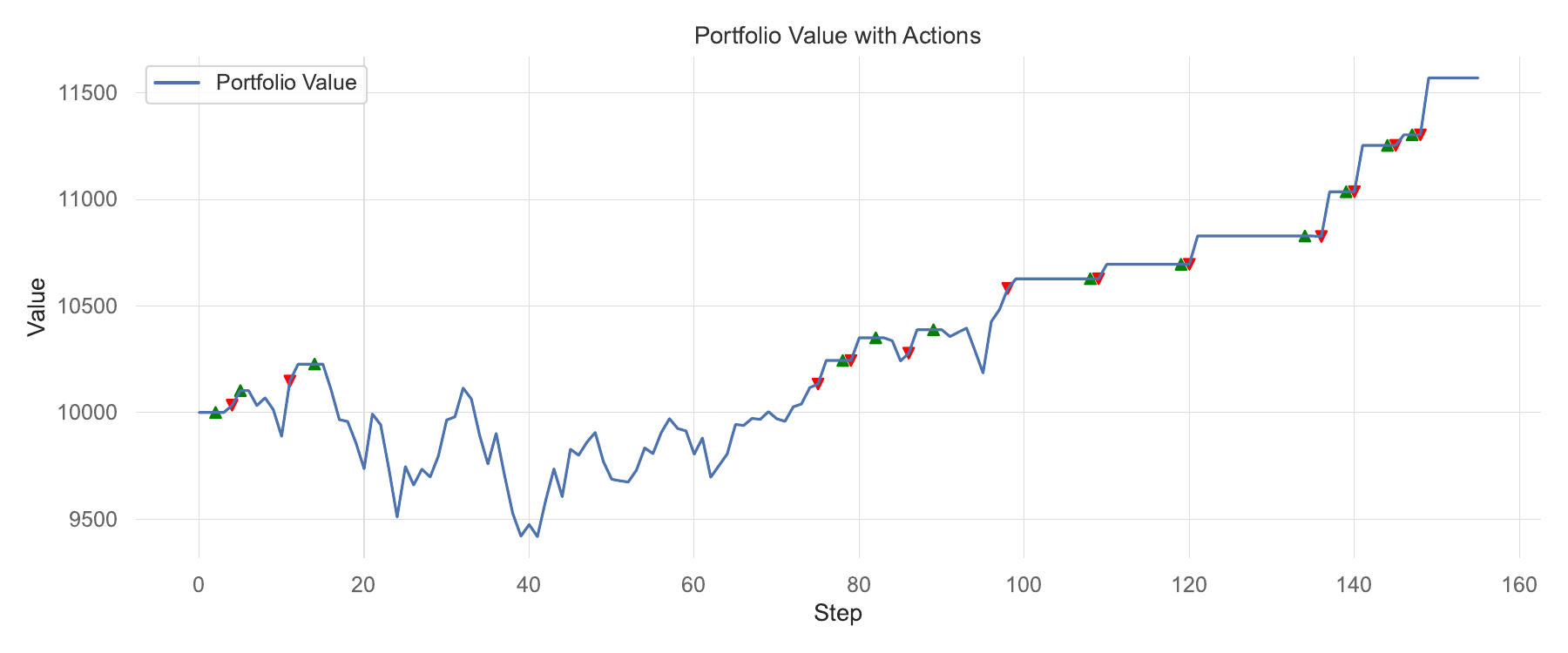}
            \caption{Quantum A3C without forecasting}
        \end{subfigure}

        \vspace{0.3cm}

        \begin{subfigure}{\linewidth}
            \includegraphics[width=\linewidth]{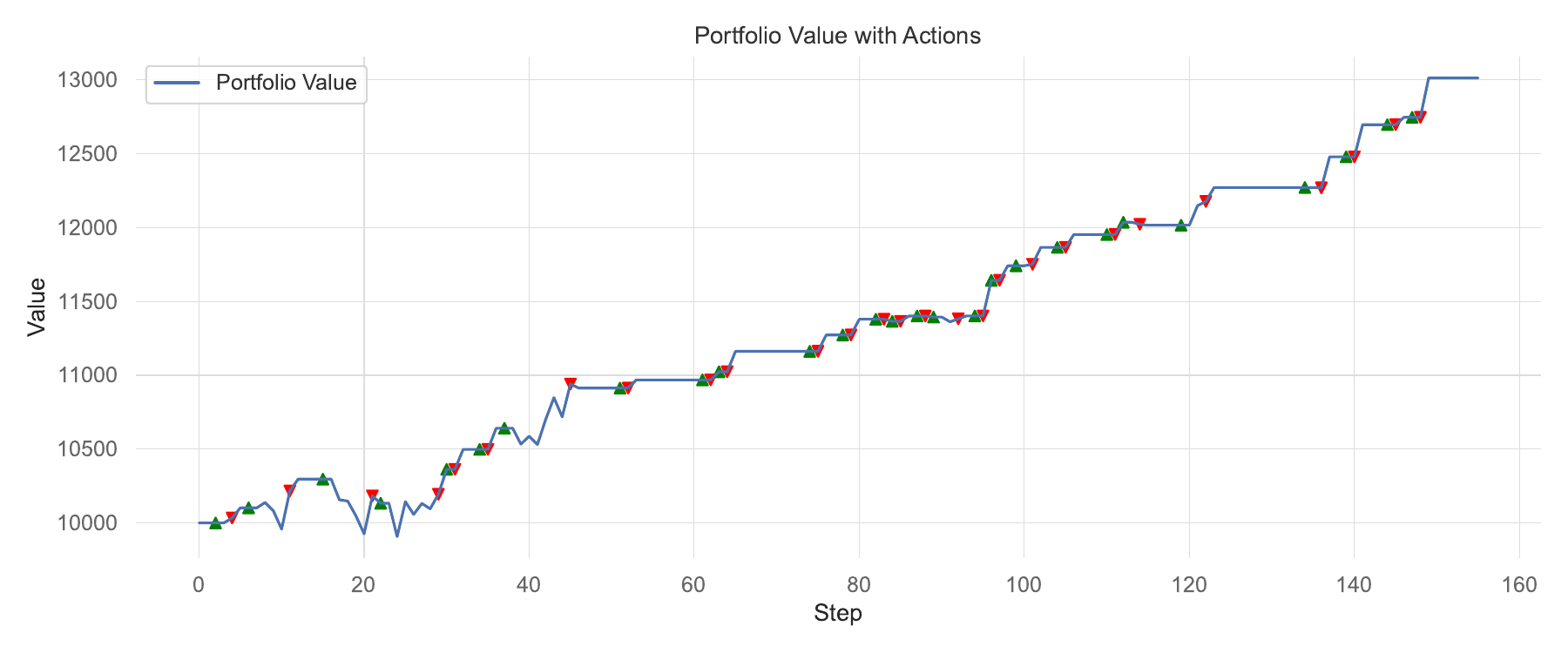}
            \caption{Quantum A3C with LSTM forecasting}
        \end{subfigure}

        \vspace{0.3cm}

        \begin{subfigure}{\linewidth}
            \includegraphics[width=\linewidth]{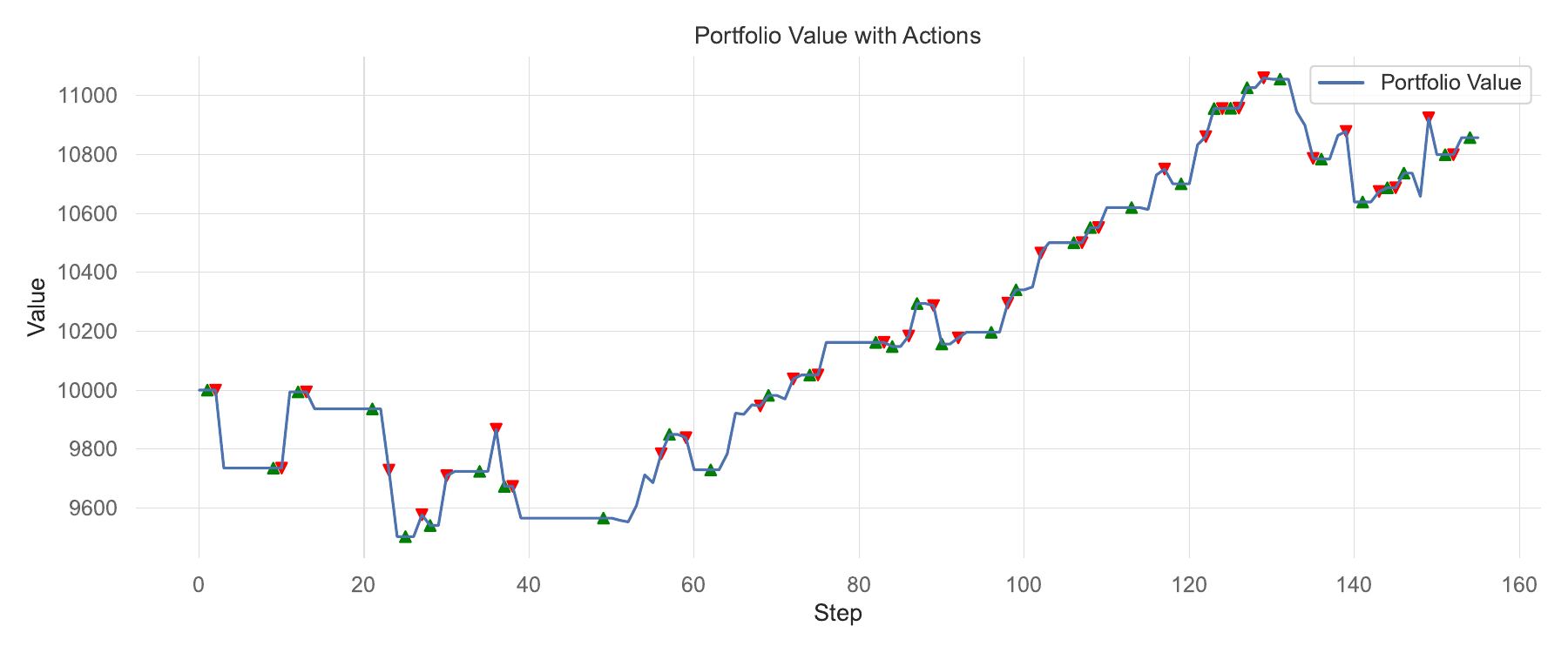}
            \caption{Random strategy baseline}
        \end{subfigure}
    \end{minipage}
    
    \caption{Left: Training curves across different reinforcement learning strategies, showing raw and average episodic rewards over 3,000 episodes. Right: Action timelines across strategies, visualizing buy/sell decisions over time for trading frequency and holding duration.}
    \label{fig:combined-curves-actions}
\end{figure*}

\subsection{Training Performance}

\begin{figure}[htbp]
\centering
\includegraphics[width=\linewidth]{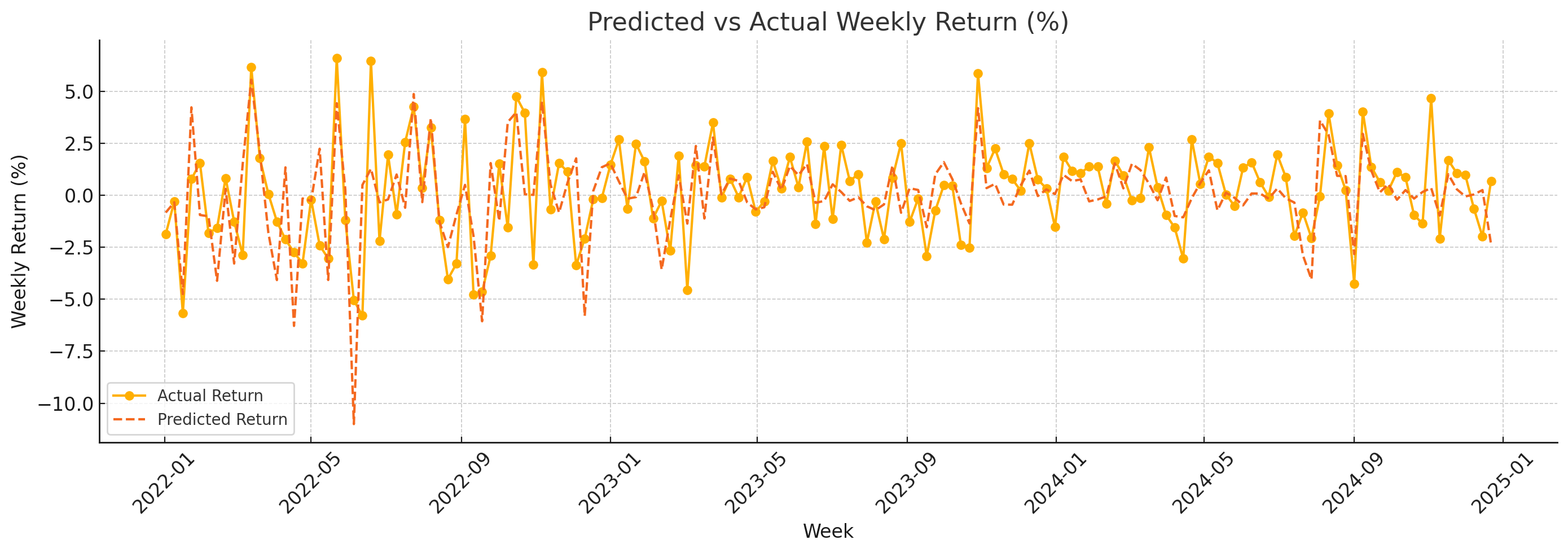}
\caption{Predicted vs Actual Weekly Return (\%) for S\&P 500 using LSTM forecasting. The model exhibits an RMSE of 2.02, a Pearson correlation of 0.595, and directional accuracy of 65.38\%, reflecting average precision in numerical values but acceptable performance in predicting market direction.}
\label{fig:lstm-prediction}
\end{figure}

 As shown in Fig.~\ref{fig:lstm-prediction}, the LSTM achieves an RMSE of 2.02, Pearson correlation of 0.595, and directional accuracy of 65.38\%. While precise value predictions are moderate, the model performs reasonably in forecasting up/down trends, providing valuable assistance to the agent.

We examine training curves (episodic rewards) across five settings: (1) Quantum A3C with LSTM signals, (2) Quantum A3C without, (3) Classical A3C with, (4) Classical A3C without, and (5) Random baseline.

Quantum A3C with LSTM exhibits the most stable and steep trajectory, indicating rapid convergence. Classical A3C models, particularly without signals, have slower convergence and higher variance. The random baseline is flat, confirming environmental consistency.

Predictive signals enhance dynamics in both quantum and classical systems, enabling more accurate anticipation of market conditions (left panel of Fig.~\ref{fig:combined-curves-actions}).

\subsection{Trading Behavior}

We analyze behaviors by counting trading cycles (buy-sell pairs). Table~\ref{tab:trading-behavior} summarizes characteristics.

\newcolumntype{C}[1]{>{\centering\arraybackslash}m{#1}}
\newcolumntype{Y}{>{\centering\arraybackslash}X}

%-------------- 表格 --------------
\begin{table*}[htbp]
\caption{Observed Trading Behavior Across Strategies}
\label{tab:trading-behavior}
\centering
\renewcommand\arraystretch{1.3}
\begin{tabular}{>{\centering\arraybackslash}m{3.5cm} 
                >{\centering\arraybackslash}m{2.5cm} 
                >{\centering\arraybackslash}m{2cm} 
                m{8cm}}
\toprule
\textbf{Strategy} & \textbf{Predict} & \textbf{Trades} & \textbf{Behavioral Characteristics} \\
\midrule
\multirow{4}{*}{Classical A3C} 
  & \multirow{2}{*}{No} 
  & \multirow{2}{*}{22} 
  & Occasional trades with longer holding periods \\
  &  &  & Most entries occurred during clear market reversals \\
  \cmidrule(lr){2-4}
  & \multirow{2}{*}{Yes} 
  & \multirow{2}{*}{10} 
  & Extremely conservative, trades only at major trend changes \\
  &  &  & Remained inactive for long periods after a few trades \\
\midrule
\multirow{4}{*}{Quantum A3C} 
  & \multirow{2}{*}{No} 
  & \multirow{2}{*}{24} 
  & Slightly more frequent than classical A3C \\
  &  &  & Reacted more aggressively to local volatility \\
  \cmidrule(lr){2-4}
  & \multirow{2}{*}{Yes} 
  & \multirow{2}{*}{54} 
  & Highest trading frequency among all \\
  &  &  & Engaged in rapid short-term trades with very short holding durations \\
\midrule
\multirow{2}{*}{Random} 
  & \multirow{2}{*}{--} 
  & \multirow{2}{*}{65} 
  & Most frequent and evenly distributed trades \\
  &  &  & No relation to price movement; trades scattered randomly \\
\bottomrule
\end{tabular}
\end{table*}

\noindent
\textbf{Frequency Ranking:}  
Random $>$ Quantum-A3C (LSTM) $>$ Quantum-A3C (no prediction) $\approx$ Classical-A3C (no prediction) $>$ Classical-A3C (LSTM)

\noindent
\textbf{Effect of Predictive Signals:}  
Predictive signals make classical A3C more conservative (22 $\rightarrow$ 10 trades), focusing on major reversals, while quantum becomes aggressive (24 $\rightarrow$ 54 trades), exploiting short-term fluctuations.

\noindent
\textbf{Holding Behavior:}  
Classical methods employ more extended periods; quantum methods are more flexible with shorter terms. Random trades are frequently made without a strategy.

\noindent
\textbf{Visualized Action Patterns:}  
The right panel of Fig.~\ref{fig:combined-curves-actions} shows timelines, revealing buy/sell distribution, frequency, holding, and trend responsiveness.

These patterns illustrate how architectural features and predictive information impact market responsiveness and decision-making.

\subsection{Trading Outcomes}

Table~\ref{tab:trading-outcomes} summarizes metrics. Quantum A3C with Predict Data yields the highest return ($\approx 30.13\%$), Sharpe ratio ($\approx 4.01$), and lowest drawdown ($\approx 3.78\%$), indicating strong performance and preservation.

\begin{table*}[ht]
\caption{Key Trading Performance Metrics Across Strategies}
\label{tab:trading-outcomes}
\centering
\renewcommand{\arraystretch}{1}
\begin{tabular*}{\textwidth}{@{\extracolsep{\fill}} l
    c
    c
    c
    c
    c
@{}}
\toprule
\textbf{Metric} & \textbf{Classical A3C} & \textbf{Classical A3C + LSTM} & \textbf{Quantum A3C} & \textbf{Quantum A3C + LSTM} & \textbf{Random} \\
\midrule
Time in Market (\%) & 53.0 & 63.0 & 59.0 & 37.0 & 46.0 \\
Cumulative Return (\%) & 14.9 & 11.5 & 15.7 & \textbf{30.13} & 8.57 \\
CAGR (\%) & 3.3 & 2.58 & 3.47 & \textbf{6.35} & 1.94 \\
Sharpe Ratio & 1.74 & 1.38 & 1.77 & \textbf{4.01} & 1.30 \\
Sortino Ratio & 2.85 & 2.15 & 2.87 & \textbf{8.76} & 1.86 \\
Smart Sharpe & 1.61 & 1.28 & 1.67 & \textbf{3.07} & 1.28 \\
Max Drawdown (\%) & -6.79 & -6.79 & -7.92 & \textbf{-3.78} & -4.97 \\
Longest Drawdown (days) & 281 & 281 & 414 & \textbf{85} & 477 \\
Volatility (Ann.) (\%) & 13.5 & 13.48 & 13.98 & \textbf{10.82} & 10.73 \\
Calmar Ratio & 0.49 & 0.38 & 0.44 & \textbf{1.68} & 0.39 \\
Gain/Pain Ratio & 0.46 & 0.32 & 0.44 & \textbf{2.05} & 0.40 \\
Profit Factor & 1.46 & 1.32 & 1.44 & \textbf{3.05} & 1.40 \\
Payoff Ratio & 1.26 & 1.09 & 1.21 & \textbf{1.65} & 0.93 \\
Tail Ratio & 1.37 & 1.18 & 1.36 & \textbf{2.61} & 0.99 \\
Omega Ratio & 1.46 & 1.32 & 1.44 & \textbf{3.05} & 1.40 \\
Ulcer Index & 0.02 & 0.02 & 0.03 & \textbf{0.01} & 0.02 \\
Recovery Factor & 2.13 & 1.69 & 1.92 & \textbf{7.07} & 1.72 \\
Serenity Index & 1.58 & 1.12 & 1.00 & \textbf{22.44} & 0.60 \\
Win Month (\%) & 53.66 & 54.64 & 54.44 & \textbf{64.91} & 60.0 \\
\bottomrule
\end{tabular*}
\end{table*}

Classical A3C with Predict Data is active (63\% exposure) but yields a lower return ($\approx 11.5\%$) and Sharpe ratio ($\approx 1.38$). Random is the worst, with a low CAGR ($\approx 1.94\%$), high drawdown (477 days), and modest Sharpe ratio ($\approx 1.30$).

LSTM Predict Data boosts Quantum A3C but minimally or negatively affects Classical A3C, suggesting that quantum better leverages signals.

Quantum A3C with Predict Data tops Serenity (22.44), Ulcer (0.01), and Gain/Pain (2.05), showing reliability in volatility. Overall, quantum RL with signals offers high returns and low risk.

\begin{figure}[htbp]
    \centering
    % --- Classical A3C ---
    \begin{subfigure}{0.48\linewidth}
        \includegraphics[width=\linewidth]{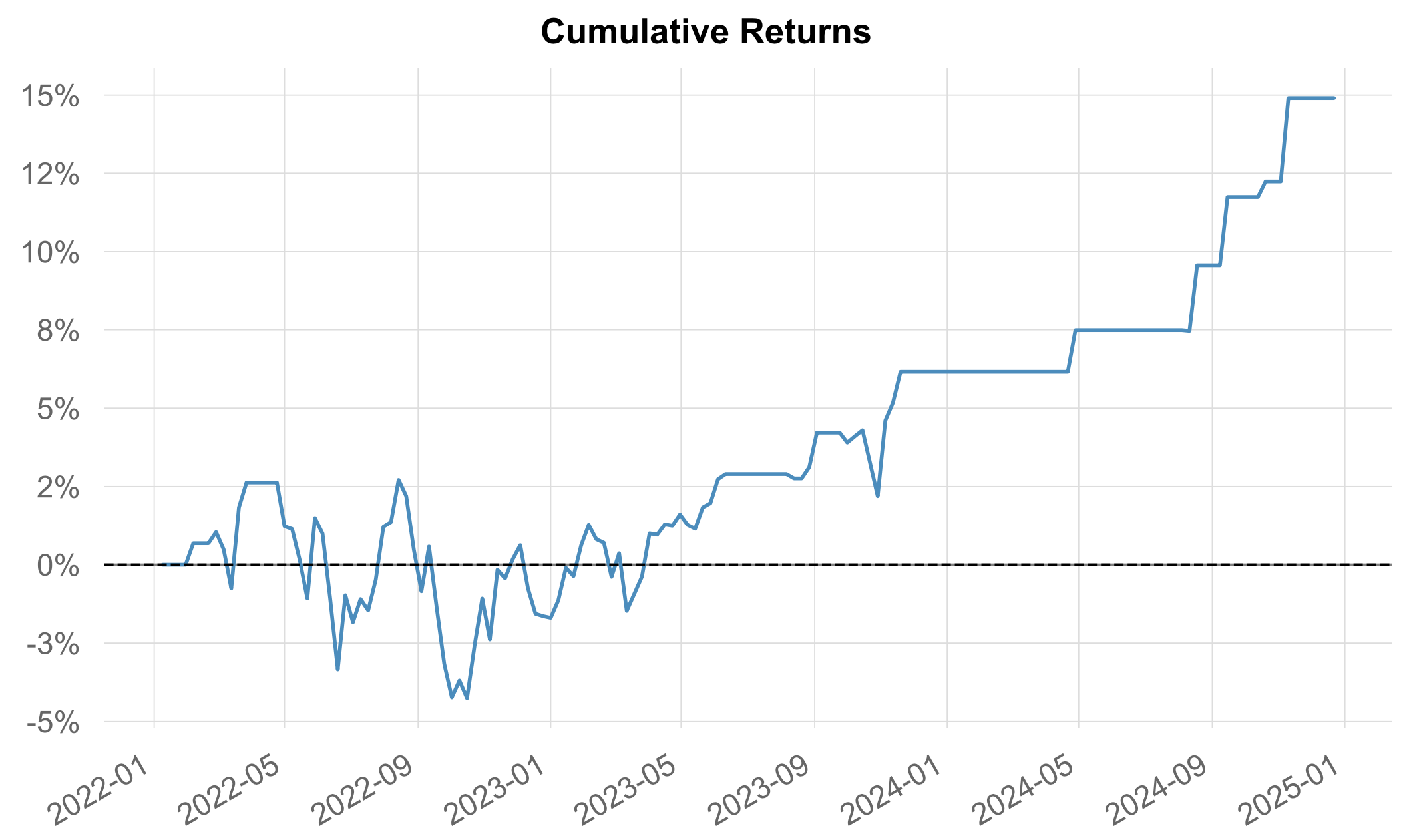}
        \caption{Classical A3C without forecasting}
    \end{subfigure}
    \hfill
    \begin{subfigure}{0.48\linewidth}
        \includegraphics[width=\linewidth]{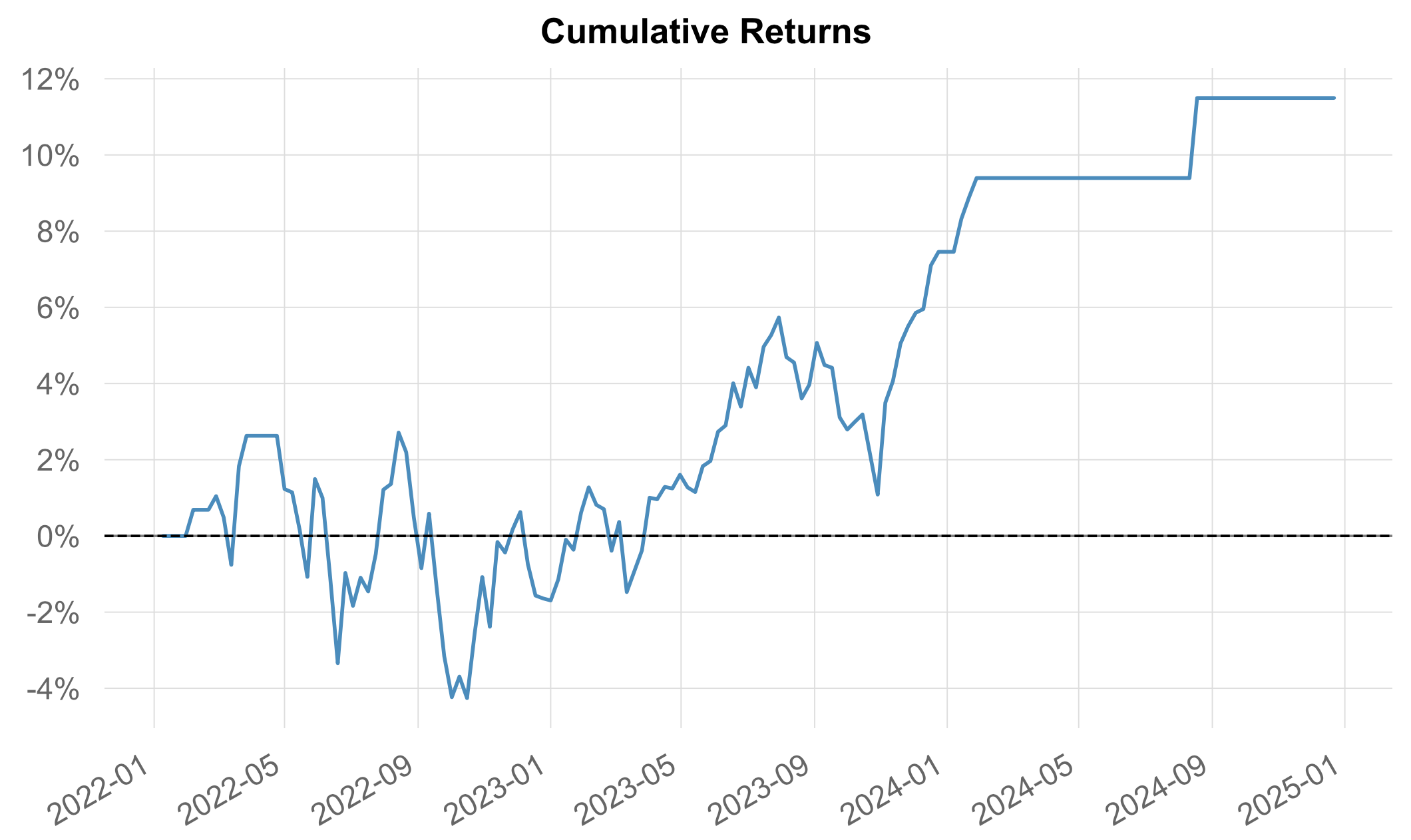}
        \caption{Classical A3C with forecasting}
    \end{subfigure}

    \vspace{0.3cm}

    % --- Quantum A3C ---
    \begin{subfigure}{0.48\linewidth}
        \includegraphics[width=\linewidth]{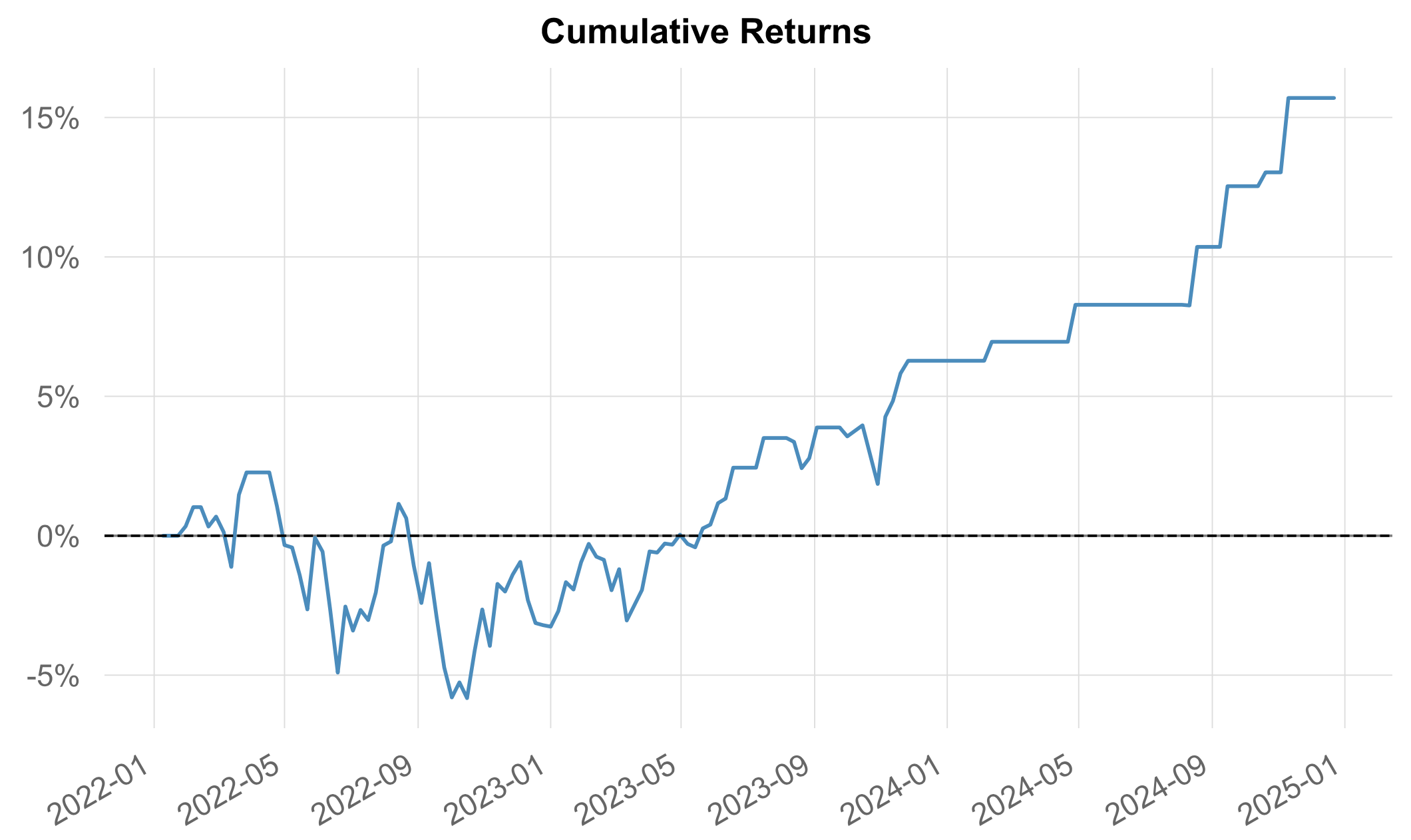}
        \caption{Quantum A3C without forecasting}
    \end{subfigure}
    \hfill
    \begin{subfigure}{0.48\linewidth}
        \includegraphics[width=\linewidth]{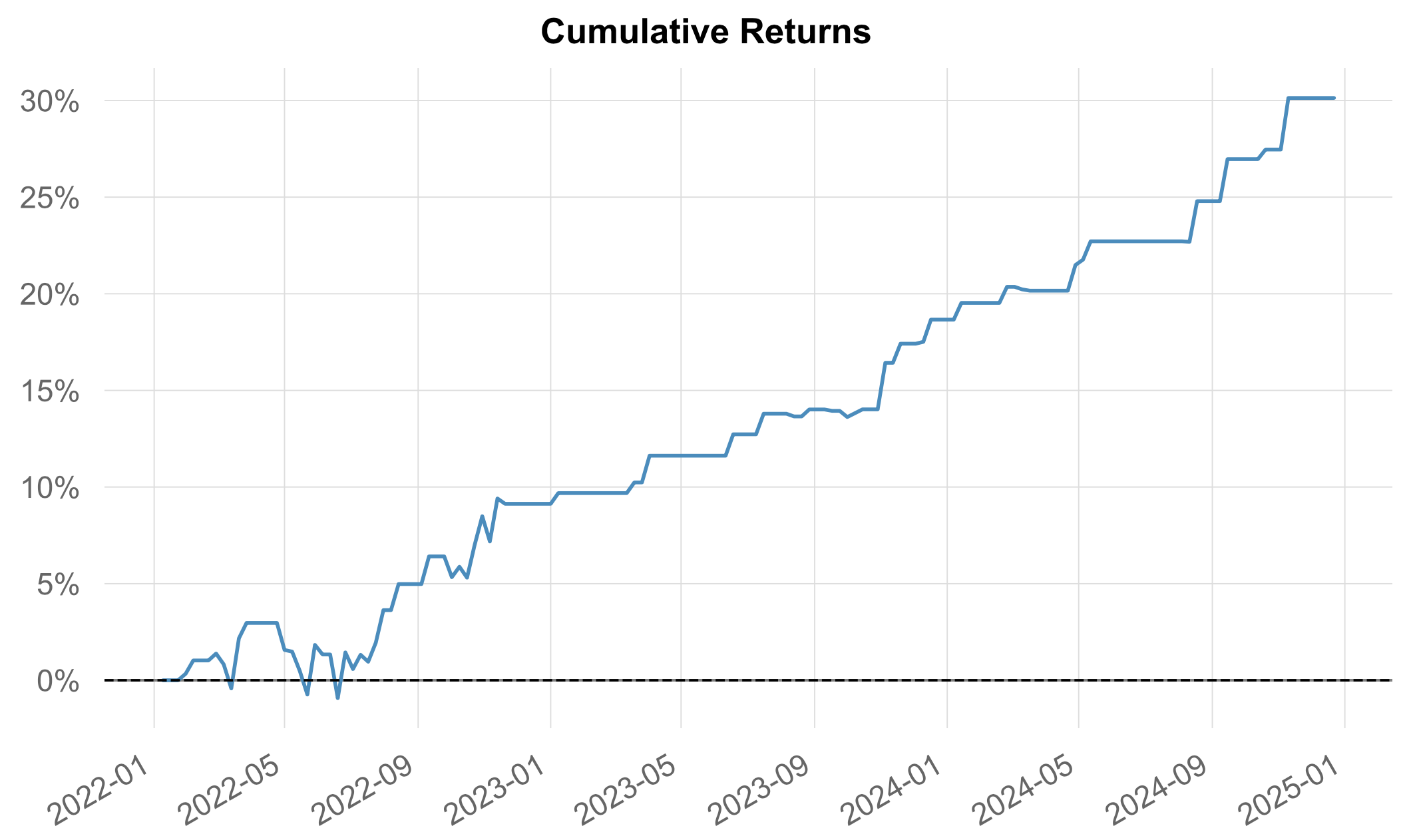}
        \caption{Quantum A3C with forecasting}
    \end{subfigure}

    \vspace{0.3cm}

    % --- Random baseline ---
    \begin{subfigure}{0.48\linewidth}
        \includegraphics[width=\linewidth]{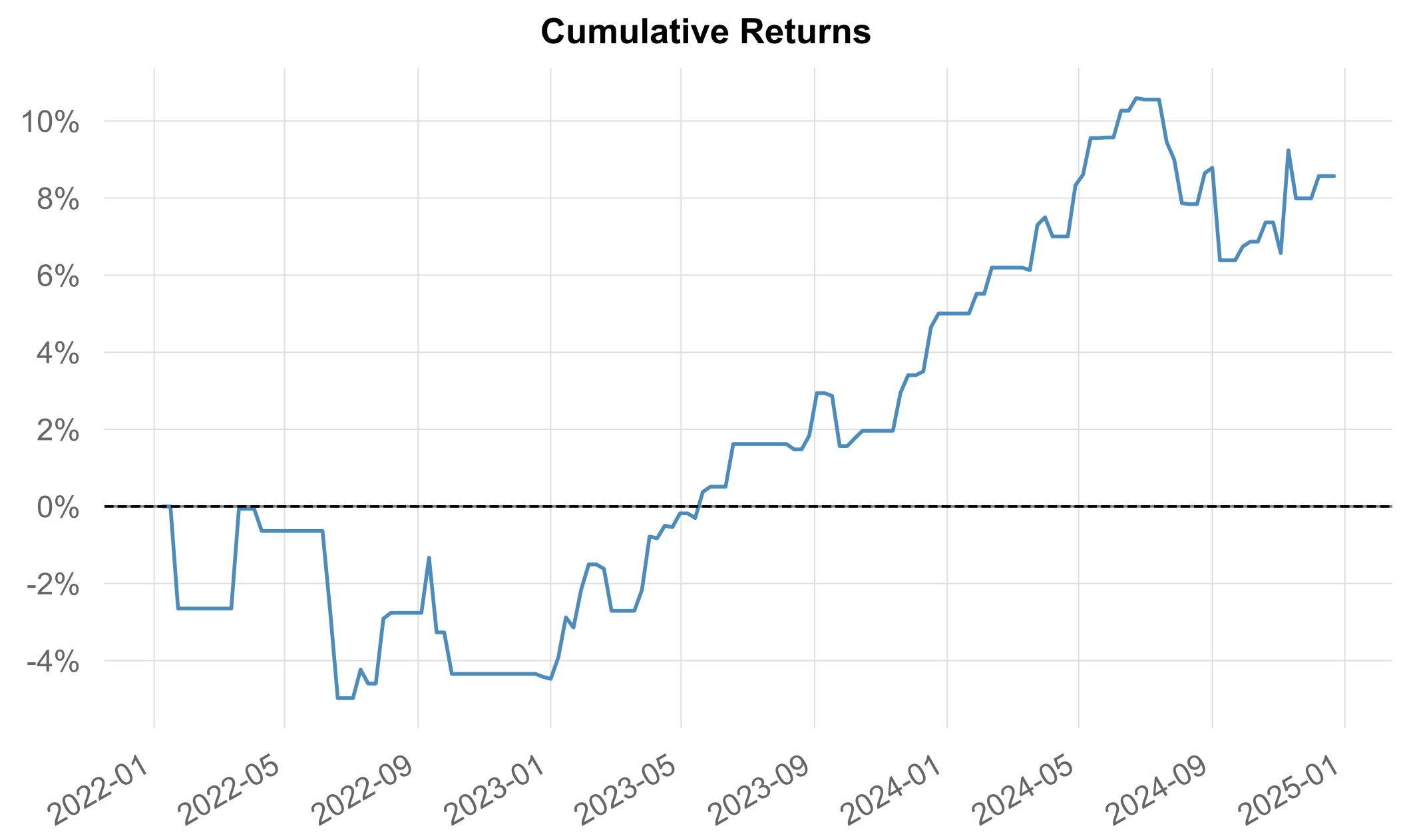}
        \caption{Random strategy baseline}
    \end{subfigure}

    \caption{Cumulative returns over the trading period for each strategy. These plots highlight return growth stability, volatility, and drawdown characteristics.}
    \label{fig:cumulative-return}
\end{figure}

\noindent
\textbf{Cumulative Return Trajectories:}  
Figure~\ref{fig:cumulative-return} visualizes returns. Quantum A3C with Predict Data exhibits a stable upward trend and minimal drawdowns, whereas Classical is erratic and Random is inconsistent.

\section{Discussion}
We compare traditional and quantum reinforcement learning (RL) agents in financial trading, both with and without predictive signals, to reveal key findings. Quantum agents, particularly those utilizing LSTM forecasting, outperform traditional ones across various metrics, including the Sharpe, Calmar, and Serenity indices, indicating superior returns and risk management capabilities. Quantum strategies effectively capture short-term dynamics, minimizing drawdowns, thanks to circuits' expressiveness for richer representations and decision boundaries in volatile markets. Predictive signals affect agents variably. For QRL, LSTM enhances decision-making by providing directional accuracy (65.38\%), enabling better anticipation of trends and handling temporal uncertainty, thereby improving training and performance. For traditional A3C, signals may hinder trading frequency due to overfitting or weak insight extraction, underscoring the need for architecture-auxiliary data alignment. Conventional methods may require attention mechanisms. Our study faces limitations, including a single-asset, limited-period focus that requires broader evaluations for generalizability. The mechanisms of quantum advantages also need theoretical scrutiny. Future work should enhance interpretability for fintech transparency and integrate quantum computing into LSTMs for improved comparisons.

\section{Conclusion}
We demonstrate that integrating variational quantum circuits into the Asynchronous Advantage Actor–Critic framework, combined with one-week-ahead LSTM macroeconomic forecasts, creates a powerful hybrid quantum–classical agent for S\&P 500 trading. By replacing the traditional feedforward encoder with a VQC, the Quantum A3C (QRL) agent captures richer, nonlinear market dynamics, achieving faster and more stable learning than classical A3C. Our novel LSTM-based predictive signals serve as an innovative environmental indicator, reducing training variance and boosting returns, with the most significant benefits derived from the synergy of quantum encoding and forecasting. In backtests on weekly data from January 2022 to April 2024, QRL with predictions achieved the highest cumulative return (approximately 30.1\%), the highest Sharpe ratio (approximately 4.01), and the lowest drawdown (approximately 3.78\%) among the methods. These results highlight QRL's contributions to enhanced risk-adjusted performance in realistic financial environments, paving the way for advanced quantum designs, real-time adaptation, and on-hardware implementation.

\bibliographystyle{IEEEtran}
\bibliography{bib/references}

\begin{thebibliography}{10}
\providecommand{\url}[1]{#1}
\csname url@samestyle\endcsname
\providecommand{\newblock}{\relax}
\providecommand{\bibinfo}[2]{#2}
\providecommand{\BIBentrySTDinterwordspacing}{\spaceskip=0pt\relax}
\providecommand{\BIBentryALTinterwordstretchfactor}{4}
\providecommand{\BIBentryALTinterwordspacing}{\spaceskip=\fontdimen2\font plus
\BIBentryALTinterwordstretchfactor\fontdimen3\font minus \fontdimen4\font\relax}
\providecommand{\BIBforeignlanguage}[2]{{%
\expandafter\ifx\csname l@#1\endcsname\relax
\typeout{** WARNING: IEEEtran.bst: No hyphenation pattern has been}%
\typeout{** loaded for the language `#1'. Using the pattern for}%
\typeout{** the default language instead.}%
\else
\language=\csname l@#1\endcsname
\fi
#2}}
\providecommand{\BIBdecl}{\relax}
\BIBdecl

\bibitem{cont2001empirical}
R.~Cont, ``Empirical properties of asset returns: stylized facts and statistical issues,'' \emph{Quantitative finance}, vol.~1, no.~2, p. 223, 2001.

\bibitem{zhang2019deep}
\BIBentryALTinterwordspacing
M.~Fazli, M.~Lashkari, H.~Taherkhani, and J.~Habibi, ``A novel experts advice aggregation framework using deep reinforcement learning for portfolio management,'' 2022. [Online]. Available: \url{https://arxiv.org/abs/2212.14477}
\BIBentrySTDinterwordspacing

\bibitem{moody1998performance}
J.~Moody, L.~Wu, Y.~Liao, and M.~Saffell, ``Performance functions and reinforcement learning for trading systems and portfolios,'' \emph{Journal of forecasting}, vol.~17, no. 5-6, pp. 441--470, 1998.

\bibitem{deng2016deep}
Y.~Deng, F.~Bao, Y.~Kong, Z.~Ren, and Q.~Dai, ``Deep direct reinforcement learning for financial signal representation and trading,'' \emph{IEEE transactions on neural networks and learning systems}, vol.~28, no.~3, pp. 653--664, 2016.

\bibitem{mnih2016asynchronous}
V.~Mnih, A.~P. Badia, M.~Mirza, A.~Graves, T.~Lillicrap, T.~Harley, D.~Silver, and K.~Kavukcuoglu, ``Asynchronous methods for deep reinforcement learning,'' in \emph{International conference on machine learning}.\hskip 1em plus 0.5em minus 0.4em\relax PmLR, 2016, pp. 1928--1937.

\bibitem{henderson2018deep}
P.~Henderson, R.~Islam, P.~Bachman, J.~Pineau, D.~Precup, and D.~Meger, ``Deep reinforcement learning that matters,'' in \emph{Proceedings of the AAAI conference on artificial intelligence}, vol.~32, no.~1, 2018.

\bibitem{biamonte2017quantum}
J.~Biamonte, P.~Wittek, N.~Pancotti, P.~Rebentrost, N.~Wiebe, and S.~Lloyd, ``Quantum machine learning,'' \emph{Nature}, vol. 549, no. 7671, pp. 195--202, 2017.

\bibitem{Schuld2015}
M.~Schuld and F.~Petruccione, ``Supervised learning with quantum computers,'' \emph{Quantum science and technology}, vol.~17, 2018.

\bibitem{Cerezo2021}
M.~Cerezo, A.~Arrasmith, R.~Babbush, S.~C. Benjamin, S.~Endo, K.~Fujii, J.~R. McClean, K.~Mitarai, X.~Yuan, L.~Cincio \emph{et~al.}, ``Variational quantum algorithms,'' \emph{Nature Reviews Physics}, vol.~3, no.~9, pp. 625--644, 2021.

\bibitem{Havlicek2019}
V.~Havl{\'\i}{\v{c}}ek, A.~D. C{\'o}rcoles, K.~Temme, A.~W. Harrow, A.~Kandala, J.~M. Chow, and J.~M. Gambetta, ``Supervised learning with quantum-enhanced feature spaces,'' \emph{Nature}, vol. 567, no. 7747, pp. 209--212, 2019.

\bibitem{tsai2025quantumfeatureoptimizationenhanced}
\BIBentryALTinterwordspacing
Y.-C. Tsai and S.~Y.-C. Chen, ``Quantum feature optimization for enhanced clustering of blockchain transaction data,'' 2025. [Online]. Available: \url{https://arxiv.org/abs/2505.16672}
\BIBentrySTDinterwordspacing

\bibitem{Dunjko2018}
V.~Dunjko, J.~M. Taylor, and H.~J. Briegel, ``Quantum-enhanced machine learning,'' \emph{Physical review letters}, vol. 117, no.~13, p. 130501, 2016.

\bibitem{Preskill2018}
J.~Preskill, ``Quantum computing in the nisq era and beyond,'' \emph{Quantum}, vol.~2, p.~79, 2018.

\bibitem{schuld2019quantum}
M.~Schuld and N.~Killoran, ``Quantum machine learning in feature hilbert spaces,'' \emph{Physical review letters}, vol. 122, no.~4, p. 040504, 2019.

\bibitem{havlivcek2019supervised}
V.~Havl{\'\i}{\v{c}}ek, A.~D. C{\'o}rcoles, K.~Temme, A.~W. Harrow, A.~Kandala, J.~M. Chow, and J.~M. Gambetta, ``Supervised learning with quantum-enhanced feature spaces,'' \emph{Nature}, vol. 567, no. 7747, pp. 209--212, 2019.

\bibitem{chen2023asynchronous}
S.~Y.-C. Chen, ``Asynchronous training of quantum reinforcement learning,'' \emph{Procedia Computer Science}, vol. 222, pp. 321--330, 2023.

\bibitem{chen2024efficient}
------, ``Efficient quantum recurrent reinforcement learning via quantum reservoir computing,'' in \emph{ICASSP 2024-2024 IEEE International Conference on Acoustics, Speech and Signal Processing (ICASSP)}.\hskip 1em plus 0.5em minus 0.4em\relax IEEE, 2024, pp. 13\,186--13\,190.

\bibitem{chen2024differentiable}
------, ``Differentiable quantum architecture search in asynchronous quantum reinforcement learning,'' in \emph{Proceedings of the 2024 IEEE International Conference on Quantum Computing and Engineering (QCE)}.\hskip 1em plus 0.5em minus 0.4em\relax IEEE, 2024, pp. 1516--1524.

\end{thebibliography}

\end{document}